\documentclass[12pt]{article} \usepackage{graphicx}
\begin{document} \title{Distribution of equilibrium edge currents}
\author{M. V. Entin, L. I. Magarill, and M. M. Mahmoodian\\\\Institute
of Semiconductor Physics SB RAS, Novosibirsk, Russia}
\date{\empty}
 \maketitle
\begin{abstract} We have studied the
distribution of equilibrium edge current density in 2D system in a strong
(quantizing) magnetic field. The case of half plane in normal magnetic
field has been considered. The transition from classical strong magnetic
field to ultraquantum limit has been investigated. We have shown that the
edge current density oscillates and decays with distance from the edge.
The oscillations have been attributed to the Fermi wavelength of
electrons. The additional component of the current smoothly depending on
the distance but sensitive to the occupation of Landau levels has been
found. The temperature suppression of oscillations has been studied.
\end{abstract}

The magnetic field acting on the low dimensional system is traditionally
considered as homogeneous and coinciding with the external field.
Nevertheless, the magnetization of such quantum system as atom
substantially changes the magnetic field in the vicinity of nucleus. This
is essential for some phenomena, e.g., NMR. The distribution of magnetic
field in the systems with spatial quantization have been studied by
authors in papers \cite{we} using the linear approximation in an external
field. This problem is close to the problem of orbital magnetism
intensively investigated in different systems with separable and
non-separable variables \cite{Az,review,Shap}. It was shown, that the
susceptibility of a large system at low temperature strongly fluctuates
and changes the sign when the Fermi level moves between the energy levels
of the system.

The goal of the present paper is to study the spatial distribution
of equilibrium currents in 2D semi-infinite plane sample subjected
to a strong magnetic field. Such mathematical setting is
applicable for a sample of arbitrary shape if its characteristic
dimensions exceed the inverse cyclotron diameter.

\subsection*{Edge current in 2D system in a finite magnetic field.}
Let us consider a semi-infinite plane $x>0,~~-\infty<y<\infty $, with hard
border in a magnetic field $B_z=B$. Assume the vector potential has gauge
$A_y=Bx$. The states in the presence of magnetic field can be described by
the longitudinal momentum $p$ and the transversal number $n$:
$\psi_{n,~p}(x)e^{i py}$. Here and below $\hbar=1$. The wave functions can
be expressed via the parabolic cylinder function $D_\nu(x)$:
\begin{eqnarray}\label{a10}
\psi_{n,p}(x)=CD_{\nu_n}\left(\sqrt{2}\left(x-x_p\right)/a\right),~~~~~
C^2\int\limits_{-x_p}^\infty
D_{\nu_n}\left(\sqrt{2}x/a\right)dx=1.
\end{eqnarray} Here
$a=\sqrt{c/eB}$ is the magnetic length, $x_0=-pa^2$. The boundary
condition $D_{\nu_n}(-\sqrt{2}pa)=0$ determines the energy levels
$E_{n,p}=\omega(\nu_n(p)+1/2)$, $n=0,1...$; $\omega=eB/m_ec$ is
the cyclotron frequency, $m_e$ is the electron effective mass.

The density of current has the form
\begin{eqnarray}\label{a11} j_y(x)=\frac{j_0}{2\pi}
\sum_{n=1}^\infty\int\limits_{-\infty}^\infty dpa
[\left(x_p-x\right)\psi_{n,p}^2(x) \bigl(f_++f_-\bigr) \nonumber\\
+\frac{gm_e}{2m_0}\bigl(f_--f_+\bigr)
\frac{a^2\partial\psi_{n,p}^2(x)}{\partial x}],
\end{eqnarray}
where $j_0=e\omega/a=e/(m_ea^3)$ is the characteristic current density,
$f_\pm=f(E_n(p)\pm g\mu_BB/2)$, \ \ $f(E)=(\exp{((E-\mu)/T)}+1)^{-1}$ is
the Fermi distribution function ($\mu,\ T$ are the chemical potential and
temperature, correspondingly), $g$ is the $g$-factor. The expression (2)
contains two contributions caused by the orbital and spin parts of the
current (the first and the second lines, correspondingly). Below we shall
neglect the spin term, assuming the smallness of the $g$-factor.

Plots of the edge current calculated according to Eqs.(\ref{a10}) and
(\ref{a11}) are presented in Figs.1-3. Let us discuss  behavior of the
edge current at low temperature. The direction of the current is
determined by the vector product of the normal and the magnetic field.
However, it does not mean the constancy of the sign of the surface current
density. Really, let only one Landau level be occupied. Consider the
states localized far from the boundary. These states are not perturbed by
the wall. The contribution to the current density from one state with
given $p$ is antisymmetric relative to the point $x_p$. As the state
approaches to the boundary the level goes up. Occupied levels (they lie
under Fermi level) have close wave functions. Averaging over momentum
compensates the current density far from the point of intersection the
Fermi level with Landau level, while near this point noncompensated
contribution of constant sign (positive in this instance) remains. When
the Fermi level approaches the lowest Landau level, the distance of the
edge state from the border grows as $\ln^{1/2}(\mu/\omega-1/2)$ and the
edge state becomes more and more ideal. Under increase of Fermi level the
wave function begins to distort due to boundary, positive part of the
contribution to the average current density is suppressed but the negative
contribution appears and the graph of current density acquires negative
minimum.

Crossing next Landau levels results in new contributions to the current
density. First these contributions are situated on large distance from the
boundary, then they move to it and merge with contributions from lower
lying states. The upper is the state the wider is the region it occupies
and the more oscillations it contributes to the current density. When we
have many occupied Landau levels, the contributions from different levels
merge to produce the Friedel oscillations of surface current density
studied by authors earlier \cite{we}. However, the number of oscillations
is restricted (unlike earlier considered limit of weak magnetic field),
because the surface current is distributed on the thickness of the order
of cyclotron diameter $2r_c$. Within this region the current density
oscillates, on the outside it decays exponentially. The number of
oscillations is defined by the number of occupied Landau levels. In the
region $x\ll 2r_c$  the effect of magnetic field is weak and the
linear-response expression from \cite{we} is valid.

As it is seen from Fig.2 the smooth dependence on coordinate $x$ is
superimposed on the space oscillations. The total edge current $J$
experiences the alternating-sign oscillations with magnetic field. These
oscillations accord with the de-Haas-van Alphen oscillations of magnetic
moment.

Note that in the limit of weak magnetic field the smooth contribution
transforms to studied in \cite{we} contribution depending linearly on the
distance, and oscillations decaying on power law spread to infinite range
from the boundary.

Temperature suppresses the oscillations of current density (see Fig.2). It
happens more effectively on large distance from boundary. The suppression
takes place due to temperature dephasing of electrons near the Fermi
surface on characteristic length $l_T=k_F/(2\pi m_eT)$, where  $k_F$ is
the Fermi momentum \cite{we}. At finite magnetic field the competition of
two lengths which cut down oscillations (cyclotron diameter and $l_T$)
occurs. The scattering suppress both spatial oscillations  and smooth
contribution as well.

The alternating-sign oscillations of total edge current seem at first
sight strange if to take into account, that at large Fermi energy the
total edge current should be diamagnetic and be described by the formula

$$J= -\frac{j_0a}{12\pi}.$$

However, in the limit of the large system the total quantity of edge
current $J$ is directly connected to the magnetic moment of the system by
the relation $M=JS/c$, where $S$ is the system area. At the same time, the
total moment at $T=0$ can be found using $\Omega$-potential of the system
\begin{eqnarray}\label{SI7} M=-\frac{\partial\Omega}{\partial
B}=\frac{j_0aS}{\pi c}
\sum\limits_{n=0}\left[\frac{\mu}{\omega}-\left(2n+1\right)\right]
\theta\left(\mu-\left(n+\frac12\right)\omega\right).
\end{eqnarray} This expression oscillates with $\mu$, jumping when the
Fermi energy crosses the Landau levels. In accord with this formula the
moment vanishes when the Fermi energy lies in the middle between the Landau
levels.

On the other hand, the edge currents in nonequilibrium conditions are
connected with the quantization of microcontact resistance in adiabatic
transport regime of quantum Hall effect. The total current in the state
$(n,p)$ is given by expression
\begin{eqnarray}\label{SI2} j_{n,p}=-e\frac{\partial
E_{n,p}}{\partial p}. \end{eqnarray} As the states are localized on $x$,
the edge current is determined by the states with $x_p$, lying near to the
border. The partial current density $j_{n,p}$ exponentially decreases with
increase of $x_p$. Therefore, summing currents over all states near the
border we obtain finite quantity. Such reasons determine the quantization
of microcontact resistance.

The total edge current in semi-infinite problem at $T=0$ looks like
\begin{eqnarray}\label{SI3}
J=-e\sum\limits_{n,p}\frac{\partial E_{n,p}}{\partial
p}~\theta\left(\mu-E_n(p)\right)=
-\frac{j_0a}{\pi}\sum\limits_{n=0}\left[\left(n+\frac12\right)-\frac{\mu}{\omega}\right]
\theta\left(\mu-\left(n+\frac12\right)\omega\right).
\end{eqnarray} The current through the structure is expressed via the difference of the
edge currents (\ref{SI3}), corresponding to chemical potentials of edges
$\mu_1$ and $\mu_2$. The difference $\mu_1-\mu_2$ in a nonequilibrium
problem coincides with the potential difference $V$, applied to the
microstructure. From here we conclude, that the nonequilibrium current is

$$J=e^2VN/h.$$

The chemical potentials of edges are equal in equilibrium, but the
edge current from one edge should be determined by the same
expression, as in non-equilibrium. However, such understanding of
the Eq.(\ref{SI3}) contradicts Eq.(\ref{SI7}) for the magnetic
moment and numerical calculations.

To find out the nature of the discrepancy, we shall find the contribution
to current density for the boundless problem from Landau levels with a
momentum lying between $p_1$ and $p_2$, where $\vert
x_{p_1}-x_{p_2}\vert\gg a$
\begin{eqnarray}\label{SI4} j_y(p_1,p_2;x)=-\frac{j_0}{\pi
}\sum\limits_n\int\limits_{p_1}^{p_2}\left(x-x_p\right)
\varphi_n^2\left(\frac{x-x_p}{a}\right)dp=-\frac{j_0}{\pi
}\sum\limits_n(\int\limits_{-\infty}^{p_2}dp...-\int\limits_{-\infty}^{p_1}dp...).
\end{eqnarray} Here $\varphi_n(\xi)$ are dimensionless normalized functions of the
harmonic oscillator. Contributions from regions close to the points
$x_{p_1}$ and $x_{p_2}$ are independent and separated in the space. Two
edge currents are presented in Eq.(\ref{SI4})(and can be isolated), but
cancel each other in the total current. The resulting contribution to a
total edge current from current density near the point $x_{p_1}$ is
located far from border and should be subtracted from a total edge
current.

This part of a current is determined by integration of $j_y(p_1,p_2;x)$
with respect to $x$ near the point $x_p$
\begin{eqnarray}\label{SI5}
\int\limits_{x_p-\Delta}^{x_p+\Delta}j_y(p_1,p_2;x)dx=
\frac{j_0a}{\pi}\sum\limits_{n=0}^{\infty}\left(n+\frac12\right)
\theta\left(\mu-\left(n+\frac12\right)\omega\right).
\end{eqnarray} \noindent Here $\Delta\gg a\sqrt{N}$. Subtracting (\ref{SI5})
from (\ref{SI3}) we obtain the correct expression for a total edge
current $J=Mc/S$, with $M$ from the (\ref{SI7}). Numerical
calculations (see Fig. 3) accord with this formula.

\subsection*{Conclusions.}

The presence of edge current results in weak change of a magnetic field
applying to the 2D system. The non-uniformly distributed current creates
the non-uniformly distributed magnetic field.

Note that the spatial inhomogeneity of a magnetic field can effect on any
responses sharply dependent on a magnetic field, in particular, on the
Shubnikov oscillations, geometrical resonances or magnetic focusing. One
can expect that the inhomogeneity will result in washout of sharp
singularities in these quantities. This gives the way to measure the
magnetization.

Other variants can be based on sensitivity of nuclear spins to a local
magnetic field. NMR-markers, placed on the certain atomic planes, can act
as indicators of local magnetic fields. One can propose to use the
constant in time magnetic field for shift of NMR line or alternating
magnetic field for excitation of transitions. The temporal change of
inhomogeneous magnetic field can be produced by control of the electron
wave functions due to alternating gate voltage. As another way one can
suggest the periodic change of temperature of electron gas by source-drain
voltage with subsequent change of $\delta B$) caused by the temperature
suppression of the Friedel oscillations.

\begin{figure}[ht]
\includegraphics[width=15cm]{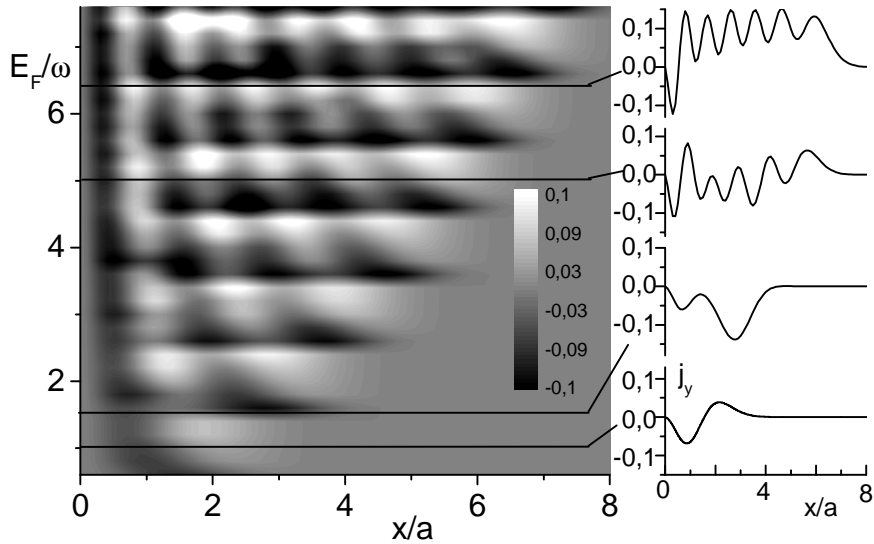}\vspace{-12cm}
\caption{On the left: relief of the edge current density in units
$j_0$ at finite magnetic field as function of the distance from
the border and Fermi energy. The grey color corresponds to zero
density of a current, white - positive, black - negative. On the
right: dependence of the current density on distance at the
selected Fermi energy (marked by straight lines in the left
figure).}\label{fig1}
\end{figure}

\begin{figure}[ht]
\includegraphics[width=15cm]{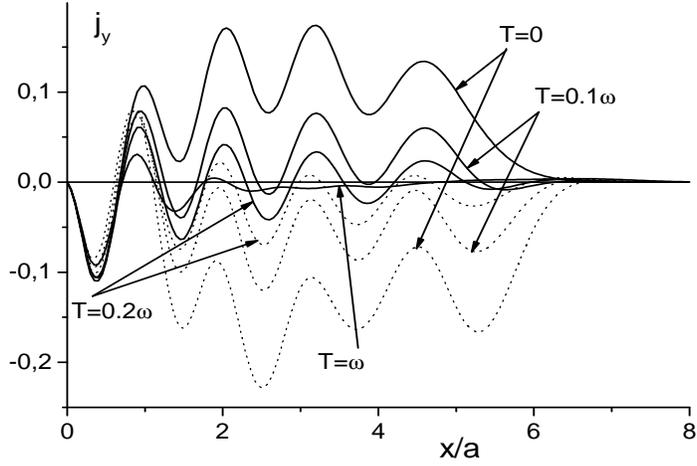}\vspace{-12cm}
\caption{Dependence of density of edge current on temperature. The
Fermi level is close to the 5th Landau level: solid curves \ - \
$E_F=4.4\omega$, dotted curves \ - \ $E_F=4.6\omega$; in these
cases the integrated currents are oppositely directed. At
temperatures $T\gg\vert E_F-4.5\omega\vert$ the curves merge. At
$T\gg\omega$ the current density ceases to depend on a magnetic
field.}\label{fig2}
\end{figure}

\begin{figure}[ht]
\includegraphics[width=15cm]{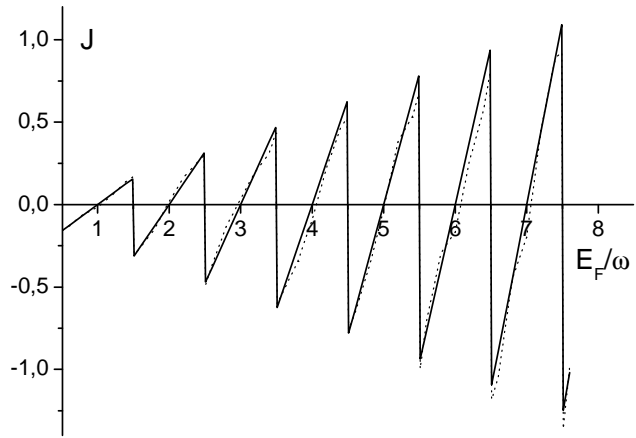}\vspace{-12cm}
\caption{The total edge current as function of a magnetic field at
$T=0$ calculated from Eq.(2) (dotted) and from Eq.(3)
(solid).}\label{fig3}
\end{figure}

\end{document}